\documentclass[12pt]{iopart}

\usepackage{graphicx}
\begin{document}

\title[Entanglement in open quantum dynamics]{Entanglement in open quantum dynamics}

\author{Aurelian Isar}

\address{National Institute of Physics and Nuclear Engineering,
P.O.Box MG-6, Bucharest-Magurele, Romania}
\ead{isar@theory.nipne.ro}
\begin{abstract}
In the framework of the theory of open systems based on completely
positive quantum dynamical semigroups,
we give a description of the continuous-variable entanglement for a system consisting of
two independent harmonic oscillators interacting with a general environment.
Using Peres-Simon necessary and sufficient criterion for separability of two-mode
Gaussian states, we describe the generation and evolution of entanglement in terms
of the covariance matrix for an arbitrary Gaussian input
state. For some values of diffusion and dissipation coefficients
describing the environment, the state keeps for all times its initial type:
separable or entangled. In other cases, entanglement generation or entanglement collapse (entanglement sudden death) take place or even a periodic collapse and revival of entanglement. We show that for certain classes of environments the initial state
evolves asymptotically to an entangled equilibrium bipartite state, while for other
values of the coefficients describing the environment, the asymptotic state is
separable. We calculate also the logarithmic negativity characterizing the degree of
entanglement of the asymptotic state.

\end{abstract}


\section{Introduction}

When two systems are immersed in an environment, then, besides
and at the same time with the quantum decoherence phenomenon, the
environment can also generate a quantum entanglement of the
two systems \cite{ben2,ben3}. In certain circumstances, the environment
enhances the entanglement and in others it suppresses the entanglement
and the state describing the two systems becomes separable. The structure of the
environment may be such that not only
the two systems become entangled, but also such that the entanglement
is maintained for a definite time or a certain amount of
entanglement survives in the asymptotic long-time regime.

In the present paper we investigate, in the framework of the theory of open systems
based on completely
positive quantum dynamical semigroups, the dynamics of the continuous-variable
entanglement for a subsystem composed of two identical harmonic
oscillators interacting with an
environment.  We are interested in discussing the correlation effect of the
environment, therefore we assume that the two systems are
independent, i.e. they do not interact directly. The initial
state of the subsystem is taken of Gaussian form and the
evolution under the quantum dynamical semigroup assures the
preservation in time of the Gaussian form of the state.

The organizing of the paper is as follows.
In Sec. 2 we write and solve the equations of motion in the Heisenberg
picture for two independent harmonic oscillators interacting with a general
environment. Then, by using the
Peres-Simon necessary and sufficient condition for separability
of two-mode Gaussian states \cite{per,sim}, we investigate in Sec. 3 the
dynamics of entanglement for the considered subsystem. In particular, with the help of the
asymptotic covariance matrix, we determine the behaviour of the entanglement in the limit
of long times. We show that for certain classes of environments
the initial state evolves asymptotically to an equilibrium
state which is entangled, while for other values of the
parameters describing the environment, the entanglement is
suppressed and the asymptotic state is separable. A summary is given in Sec. 4

\section{Equations of motion for two independent harmonic oscillators}

We study the dynamics
of the subsystem composed of two identical non-interacting
oscillators in weak interaction with an environment. In the axiomatic formalism
based on completely positive quantum dynamical semigroups, the irreversible time
evolution of an open
system is described by the following general quantum Markovian master equation
for an operator $A$ (Heisenberg representation) \cite{lin,rev}:
\begin{eqnarray}{dA(t)\over dt}={i\over \hbar}[H,A(t)]+{1\over
2\hbar}\sum_j(V_j^{\dagger}[A(t),
V_j]+[V_j^{\dagger},A(t)]V_j).\label{masteq}\end{eqnarray}
Here, $H$ denotes the Hamiltonian of the open system
and the operators $V_j, V_j^\dagger,$ defined on the Hilbert space of $H,$
represent the interaction of the open system
with the environment. Being interested
in the set of Gaussian states, we introduce such quantum
dynamical semigroups that preserve this set. Therefore $H$ is
taken to be a polynomial of second degree in the coordinates
$x,y$ and momenta $p_x,p_y$ of the two quantum oscillators and
$V_j,V_j^{\dagger}$ are taken polynomials of first degree
in these canonical observables. Then in the linear space
spanned by the coordinates and momenta there exist only four
linearly independent operators $V_{j=1,2,3,4}$ \cite{san}
: \begin{eqnarray}
V_j=a_{xj}p_x+a_{yj}p_y+b_{xj}x+b_{yj}y,\end{eqnarray} where
$a_{xj},a_{yj},b_{xj},b_{yj}\in {\bf C}.$
The Hamiltonian $H$ of the two uncoupled identical harmonic
oscillators of mass $m$ and frequency $\omega$ is given by \begin{eqnarray}
H={1\over 2m}(p_x^2+p_y^2)+{m\omega^2\over
2}(x^2+y^2).\end{eqnarray}

The fact that the evolution is given by a dynamical semigroup
implies the positivity of the following matrix formed by the
scalar products of the four vectors $ {\bi a}_x, {\bi b}_x,
{\bi a}_y, {\bi b}_y$ whose entries are the components $a_{xj},b_{xj},a_{yj},b_{yj},$
respectively:
\begin{eqnarray}\frac{1}{2} \hbar \left(\matrix
{({\bi a}_x {\bi a}_x)&({\bi a}_x {\bi b}_x) &({\bi a}_x
{\bi a}_y)&({\bi a}_x {\bi b}_y) \cr ({\bi b}_x {\bi a}_x)&({\bi
b}_x {\bi b}_x) &({\bi b}_x {\bi a}_y)&({\bi b}_x {\bi b}_y)
\cr ({\bi a}_y {\bi a}_x)&({\bi a}_y {\bi b}_x) &({\bi a}_y
{\bi a}_y)&({\bi a}_y {\bi b}_y) \cr ({\bi b}_y {\bi
a}_x)&({\bi b}_y {\bi b}_x) &({\bi b}_y {\bi a}_y)&({\bi b}_y
{\bi b}_y)}\right).
\end{eqnarray}
We take this matrix of the following form, where
all coefficients $D_{xx}, D_{xp_x},$... and $\lambda$ are real quantities (we
put, for simplicity, $\hbar=1$):
\begin{eqnarray} \left(\matrix{D_{xx}&- D_{xp_x} - i \lambda/2&D_{xy}& -
D_{xp_y} \cr - D_{xp_x} + i \lambda/2&D_{p_x p_x}&-
D_{yp_x}&D_{p_x p_y} \cr D_{xy}&- D_{y p_x}&D_{yy}&- D_{y p_y}
- i \lambda/2 \cr - D_{xp_y} &D_{p_x p_y}&- D_{yp_y} + i
\lambda/2&D_{p_y p_y}}\right).\label{coef} \end{eqnarray} It follows that
the principal minors of this matrix are positive or zero. From
the Cauchy-Schwarz inequality the following relations for the
coefficients defined in Eq. (\ref{coef}) hold: \begin{eqnarray}
D_{xx}D_{p_xp_x}-D^2_{xp_x}\ge\frac{\lambda^2}{4},~
D_{yy}D_{p_yp_y}-D^2_{yp_y}\ge\frac{\lambda^2}{4},\nonumber\\
D_{xx}D_{yy}-D^2_{xy}\ge0,~
D_{xx}D_{p_yp_y}-D^2_{xp_y}\ge 0,\nonumber \\
D_{yy}D_{p_xp_x}-D^2_{yp_x}\ge 0,~
D_{p_xp_x}D_{p_yp_y}-D^2_{p_xp_y}\ge 0.\label{coefineq}\end{eqnarray}
The matrix of the coefficients (\ref{coef}) can be conveniently
written as
\begin{eqnarray}
\left(\begin{array}{cc}C_1&C_3\\
 {C_3}^{\dagger} &C_2 \end{array}\right),\label{subm}
\end{eqnarray}
in terms of $2\times 2$ matrices
$C_1={C_1}^\dagger$, $C_2={C_2}^\dagger$ and ${C_3}$. This
decomposition has a direct physical interpretation: the
elements containing the diagonal contributions $C_1$ and $C_2$
represent diffusion and dissipation coefficients corresponding
to the first, respectively the second, system in absence of the
other, while the elements in $C_3$ represent environment
generated couplings between the two, initially independent,
oscillators.

We introduce the following $4\times 4$ bimodal covariance matrix:
\begin{eqnarray}\sigma(t)=\left(\matrix{\sigma_{xx}&\sigma_{xp_x} &\sigma_{xy}&
\sigma_{xp_y}\cr \sigma_{xp_x}&\sigma_{p_xp_x}&\sigma_{yp_x}
&\sigma_{p_xp_y}\cr \sigma_{xy}&\sigma_{yp_x}&\sigma_{yy}
&\sigma_{yp_y}\cr \sigma_{xp_y}&\sigma_{p_xp_y}&\sigma_{yp_y}
&\sigma_{p_yp_y}}\right),\label{covar} \end{eqnarray} with the correlations of operators $A_1$ and
$A_2,$ defined by using the density operator $\rho$,
describing the initial state of the quantum system, as follows:
\begin{eqnarray}\sigma_{A_1A_2}(t)\nonumber\\={1\over 2}{\rm
Tr}(\rho(A_1A_2+A_2A_1)(t))-{\rm Tr}(\rho A_1(t)){\rm Tr}(\rho A_2(t)).\end{eqnarray}
By direct calculation we obtain \cite{san} (${\rm T}$ denotes the transposed matrix):
\begin{eqnarray}{d \sigma(t)\over
dt} = Y \sigma(t) + \sigma(t) Y^{\rm T}+2 D,\label{vareq}\end{eqnarray} where
\begin{eqnarray} Y=\left(\matrix{ -\lambda&1/m&0 &0\cr -m\omega^2&-\lambda&0&
0\cr 0&0&-\lambda&1/m \cr 0&0&-m\omega^2&-\lambda}\right),\\
D=\left(\matrix{
D_{xx}& D_{xp_x} &D_{xy}& D_{xp_y} \cr D_{xp_x}&D_{p_x p_x}&
D_{yp_x}&D_{p_x p_y} \cr D_{xy}& D_{y p_x}&D_{yy}& D_{y p_y}
\cr D_{xp_y} &D_{p_x p_y}& D_{yp_y} &D_{p_y p_y}} \right).\end{eqnarray}
The time-dependent
solution of Eq. (\ref{vareq}) is given by \cite{san}
\begin{eqnarray}\sigma(t)= M(t)(\sigma(0)-\sigma(\infty)) M^{\rm
T}(t)+\sigma(\infty),\label{covart}\end{eqnarray} where the matrix $M(t)=\exp(Yt)$ has to fulfill
the condition $\lim_{t\to\infty} M(t) = 0.$
In order that this limit exists, $Y$ must only have eigenvalues
with negative real parts. The values at infinity are obtained
from the equation \cite{san} \begin{eqnarray}
Y\sigma(\infty)+\sigma(\infty) Y^{\rm T}=-2 D.\label{covarinf}\end{eqnarray}

\section{Dynamics of entanglement}

The two-mode Gaussian state is entirely specified by its
covariance matrix (\ref{covar}), which is a real,
symmetric and positive matrix with the following block
structure:
\begin{eqnarray}
\sigma(t)=\left(\begin{array}{cc}A&C\\
C^{\rm T}&B \end{array}\right),
\end{eqnarray}
where $A$, $B$ and $C$ are $2\times 2$ matrices. Their entries
are correlations of the canonical operators $x,y,p_x,p_y$, $A$
and $B$ denote the symmetric covariance matrices for the
individual reduced one-mode states, while the matrix $C$
contains the cross-correlations between modes. The elements of
the covariance matrix depend on $Y$ and $D$ and can be
calculated from Eqs. (\ref{covart}), (\ref{covarinf}). Since the two oscillators are identical, it is natural to consider environments for which the two
diagonal submatrices in Eq. (\ref{subm}) are equal, $C_1=C_2,$ and the matrix $C_3$ is symmetric,
so that in the following we take $D_{xx}=D_{yy}, D_{xp_x}=D_{yp_y},
D_{p_xp_x}=D_{p_yp_y}, D_{xp_y}=D_{yp_x}.$ Then both unimodal covariance
matrices are equal, $A=B,$ and the entanglement matrix $C$ is
symmetric.

\subsection{Asymptotic entanglement}

First we analyze the existence of entanglement in the limit of large times. With the chosen coefficients, we obtain from Eq. (\ref{covarinf}) the
following elements of the asymptotic entanglement matrix $C(\infty)$:
\begin{eqnarray}\sigma_{xy} (\infty) =
\frac{m^2(2\lambda^2+\omega^2)D_{xy}+2m\lambda
D_{xp_y}+D_{p_xp_y}} {2m^2\lambda(\lambda^2+\omega^2)},\end{eqnarray}
\begin{eqnarray}\sigma_{xp_y}(\infty)=
\sigma_{yp_x}(\infty)=\frac{-m^2\omega^2 D_{xy}+2m\lambda
D_{xp_y}+ D_{p_xp_y}}{2m(\lambda^2+\omega^2)},\end{eqnarray}
\begin{eqnarray}\sigma_{p_xp_y} (\infty) =
\frac{m^2\omega^4D_{xy}-2m\omega^2\lambda D_{xp_y}+(2\lambda^2
+\omega^2)D_{p_xp_y}}{2\lambda(\lambda^2+\omega^2)}.\end{eqnarray} The
elements of matrices $A(\infty)$ and $B(\infty)$ are obtained by putting $x=y$ in the previous
expressions. We calculate the determinant of the entanglement matrix and obtain:
\begin{eqnarray} \det
C(\infty)=\frac{1}{4\lambda^2(\lambda^2+\omega^2)}\nonumber\\\times[(m\omega^2D_{xy}+
\frac{1}{m}
D_{p_xp_y})^2+4\lambda^2(D_{xy}D_{p_xp_y}-D_{xp_y}^2)].\end{eqnarray}
It is interesting that the general theory of open quantum
systems allows couplings via the environment between uncoupled
oscillators. According to the definitions of the environment
parameters, the diffusion coefficients above can be different
from zero and therefore can simulate an interaction between the uncoupled
oscillators. Indeed, Gaussian states with $\det C\ge 0$ are
separable states, but for $\det C <0$ it may be possible that
the asymptotic equilibrium states are entangled, as will be shown next.

On general grounds, one expects that the effects of
decoherence, counteracting entanglement production, be dominant
in the long-time regime, so that no quantum correlation (entanglement) is expected to be left at infinity. Nevertheless, there are situations in which the environment allows the
presence of entangled asymptotic equilibrium states. In order
to investigate whether an external environment can actually
entangle the two independent systems, we use the partial
transposition criterion \cite{per,sim}: a state is
entangled if and only if the operation of partial transposition
does not preserve its positivity. For the particular case of Gaussian states, Simon \cite{sim} obtained the
following necessary and sufficient criterion for separability:
$S\ge 0,$ where \begin{eqnarray} S\equiv\det A \det B\nonumber\\+(\frac{1}{4} -|\det
C|)^2- {\rm Tr}[AJCJBJC^{\rm T}J]- \frac{1}{4}(\det A+\det B)
\label{sim1}\end{eqnarray} and $J$ is the $2\times 2$ symplectic matrix
\begin{eqnarray}
J=\left(\begin{array}{cc}0&1\\
-1&0\end{array}\right).
\end{eqnarray}
In the following we consider an environment characterized by diffusion coefficients of the form $m^2\omega^2D_{xx}=D_{p_xp_x},~D_{xp_x}=0,
~m^2\omega^2D_{xy}=D_{p_xp_y}.$ This corresponds to the case when the asymptotic state is a Gibbs state \cite{rev}. Then Simon
expression (\ref{sim1}) takes the following form in the limit of large times: \begin{eqnarray} S(\infty)=
\left(\frac{m^2\omega^2(D_{xx}^2-D_{xy}^2)}{\lambda^2}+
\frac{D_{xp_y}^2}{\lambda^2+\omega^2}-\frac{1}{4}\right)^2\nonumber\\-4\frac
{m^2\omega^2D_{xx}^2D_{xp_y}^2}{\lambda^2(\lambda^2+
\omega^2)}.\label{sim2}\end{eqnarray} For environments characterized by
such coefficients that the expression $S(\infty)$ (\ref{sim2}) is strictly negative,
the asymptotic final state is entangled. In particular, for
$D_{xy}=0$ we obtain that $S(\infty)<0,$ i.e. the asymptotic final
state is entangled, for the following range of values of the
coefficient $D_{xp_y}$ characterizing the environment \cite{arus,aijqi}:
\begin{eqnarray}
\frac{m\omega
D_{xx}}{\lambda}-\frac{1}{2}<\frac{D_{xp_y}}{\sqrt{\lambda^2
+\omega^2}}<\frac{m\omega
D_{xx}}{\lambda}+\frac{1}{2},\label{insep}\end{eqnarray} where the
diffusion coefficient $D_{xx}$ satisfies the condition $m\omega
D_{xx}/\lambda\ge 1/2,$ equivalent with the unimodal
uncertainty relation. We remind that, according to inequalities (\ref{coefineq}), the coefficients have to fulfill also the constraint $D_{xx}\ge D_{xp_y}.$ If the coefficients do not fulfil the
inequalities (\ref{insep}), then $S(\infty)\ge 0$ and the
asymptotic state of the considered system is
separable.

\begin{figure}
\resizebox{0.66\columnwidth}{!}
{
\includegraphics{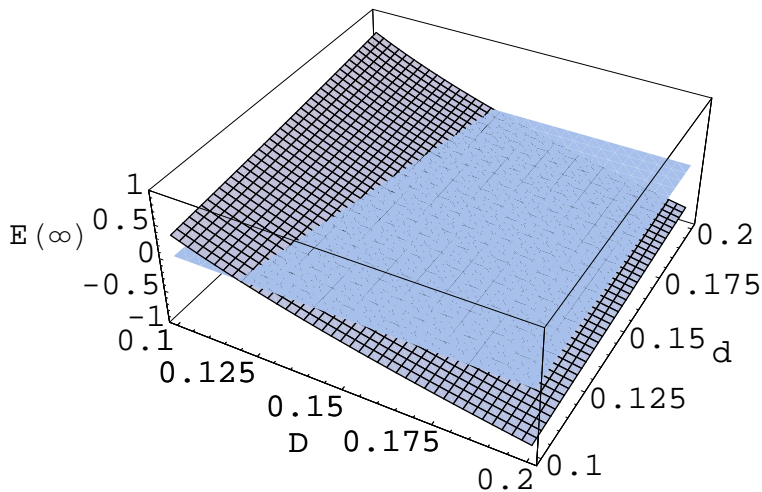}}
\caption{Asymptotic logarithmic negativity $E(\infty)$ versus environment coefficients $D$ and $d.$ We have taken $m=\omega=\hbar=1$ and $\lambda=0.2.$ According to inequalities (\ref{coefineq}), only the part of the plot corresponding to the constraint $D\ge d$ has a physical meaning.}
\label{fig:1}
\end{figure}

\subsection{Asymptotic logarithmic negativity}

We apply the measure of entanglement based on negative eigenvalues
of the partial transpose of the subsystem density matrix. For a Gaussian density
operator, the negativity is completely defined
by the symplectic spectrum of the partial transpose of the
covariance matrix. The logarithmic negativity $E=-\frac{1}{2}\log_2[4f(\sigma)]$
determines the strength of entanglement for $E>0.$ If $E\le 0,$ then the state is
separable. Here \begin{eqnarray} f(\sigma)=\frac{1}{2}(\det A +\det
B)-\det C\nonumber\\-\left({\left[\frac{1}{2}(\det A+\det B)-\det
C\right]^2-\det\sigma}\right)^{1/2}.\end{eqnarray}  In our case the asymptotic logarithmic negativity
has the form
\begin{eqnarray} E(\infty)=-\log_2\left[2\left|\frac{m\omega
D_{xx}}{\lambda}-\frac{D_{xp_y}}{\sqrt{\lambda^2
+\omega^2}}\right|\right].\end{eqnarray}
It depends only on the diffusion and dissipation coefficients
characterizing the environment and does not depend on the initial
Gaussian state. As an example, in Figure 1 we represent the asymptotic logarithmic negativity $E(\infty)$ versus diffusion coefficients $D_{xx}\equiv D$ and $D_{xp_y}\equiv d.$ We notice that for some range of these coefficients, $E(\infty)$ takes strictly positive values, measuring the degree of entanglement of the corresponding asymptotic entangled states. Again, the coefficients have to fulfill the constraint $D_{xx}\ge D_{xp_y},$ so that in Figure 1 only the part of the plot corresponding to $D\ge d$ has a physical meaning for our model.

\subsection{Time evolution of entanglement}

\begin{figure}
\resizebox{0.66\columnwidth}{!} {
\includegraphics{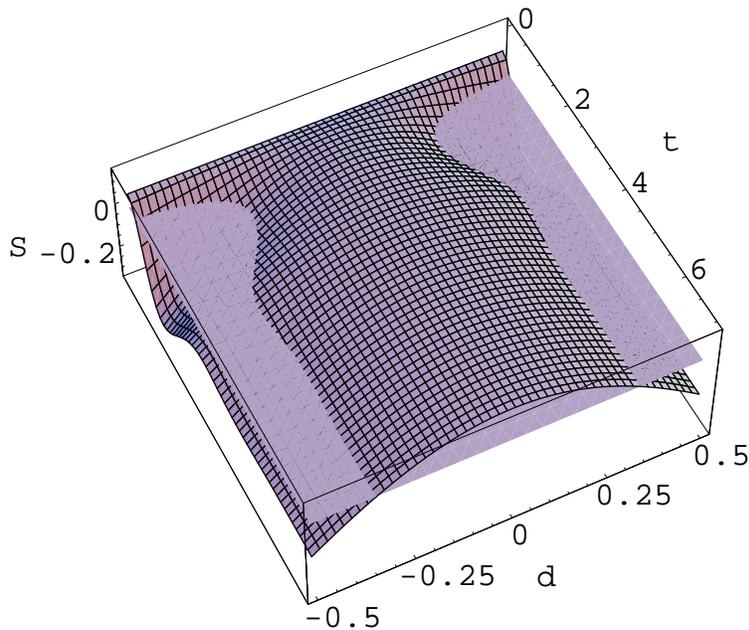}
}
\caption{Simon separability function $S$ versus time $t$
and environment coefficient $d,$ for $\lambda=0.5,$ $D=0.4$ and a separable initial Gaussian state with initial correlations $\sigma_{xx}(0)=1,~\sigma_{p_xp_x}(0)=1/2,~\sigma_{xp_x}(0)=\sigma_{xy}(0)=\sigma_{p_xp_y}(0)=~\sigma_{xp_y}(0)=0.$ We have taken $m=\omega=\hbar=1.$
}
\label{fig:2}
\end{figure}

In order to describe the dynamics of entanglement, we have to analyze the time evolution of the Simon function (\ref{sim1}). We consider separately the two cases, according to the type of the initial Gaussian state: separable or entangled.

1) To illustrate a possible generation of the entanglement, we represent in Figure 2 the function $S(t)$ versus time and diffusion coefficient  $D_{xp_y}\equiv d$ for a separable initial Gaussian state with initial correlations $\sigma_{xx}(0)=1,~\sigma_{p_xp_x}(0)=1/2,~\sigma_{xp_x}(0)=\sigma_{xy}(0)=\sigma_{p_xp_y}(0)=\sigma_{xp_y}(0)=0.$ We notice that, according to Peres-Simon criterion, for relatively small absolute values of the coefficient $d,$ the initial separable state remains separable for all times, while for larger absolute values of $d,$ at some finite moment of time the state becomes entangled. In some cases the entanglement is only temporarily generated and the state becomes again separable after a certain moment of time. In other cases, namely for even larger absolute values of the coefficient $d,$  the generated entangled state remains entangled forever, including the asymptotic final state.

2) The evolution of an entangled initial state is illustrated in Figure 3, where we represent the function $S(t)$ versus time and diffusion coefficient  $d$ for an initial entangled Gaussian state with  initial correlations $\sigma_{xx}(0)=1,~\sigma_{p_xp_x}(0)=1/2,~\sigma_{xp_x}(0)=0,~
\sigma_{xy}(0)=1/2,~\sigma_{p_xp_y}(0)=-1/2,~\sigma_{xp_y}(0)=0.$ We notice that for relatively large absolute values of the coefficient $d,$ the initial entangled state remains entangled for all times, while for smaller absolute values of $d,$ at some finite moment of time the state becomes separable. This is the well-known phenomenon of entanglement sudden death. Depending on the values of the coefficient $d$, it is also possible to have a repeated collapse and revival of the entanglement.

\begin{figure}
\resizebox{0.66\columnwidth}{!} {
\includegraphics{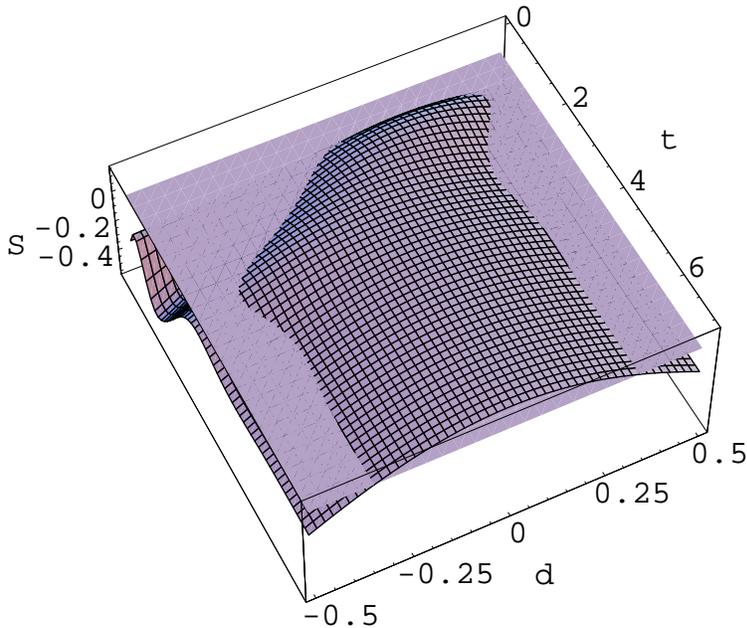}
}
\caption{Same as in Figure 2, for an entangled initial Gaussian state with initial correlations $\sigma_{xx}(0)=1,~\sigma_{p_xp_x}(0)=1/2,~\sigma_{xp_x}(0)=0,~
\sigma_{xy}(0)=1/2,~\sigma_{p_xp_y}(0)=-1/2,~\sigma_{xp_y}(0)=0.$
}
\label{fig:3}
\end{figure}

\section{Summary}

In the framework of the theory of open quantum systems based on completely
positive quantum dynamical
semigroups, we investigated
the existence of the quantum entanglement for a subsystem
composed of two uncoupled identical harmonic oscillators
interacting with a general environment.
By using the Peres-Simon
necessary and sufficient condition for separability of two-mode
Gaussian states, we have described the generation and evolution of entanglement in terms
of the covariance matrix for an arbitrary Gaussian input
state. For some values of diffusion and dissipation coefficients
describing the environment, the state keeps for all times its initial type:
separable or entangled. In other cases, entanglement generation or entanglement collapse (entanglement sudden death) take place or even one can notice a repeated collapse and revival of entanglement.
We have also shown that, independent of the type of the
initial state, for certain classes of
environments the initial state evolves asymptotically to an
equilibrium state which is entangled, while for other values of the
coefficients describing the environment, the asymptotic state
is separable. We determined also the logarithmic negativity characterizing the
degree of entanglement of the asymptotic state.
The existence of quantum correlations between the two considered harmonic oscillators interacting with a common environment is the result of the competition between entanglement and quantum decoherence.

Due to the increased interest manifested towards
the continuous-variable approach to quantum information theory, these results,
in particular the
possibility of maintaining a bipartite entanglement in a
diffusive-dissipative environment for asymptotic long
times, might be useful in controlling the entanglement in open systems and also for applications in quantum information
processing and communication.

\ack

The author acknowledges the financial support received within
the Project CEEX 68/2005.


\section*{References}
\begin{thebibliography}{9}

\bibitem{ben2} Benatti F and Floreanini R 2005 {\it Int. J. Mod.
Phys. B} {\bf 19} 3063

\bibitem{ben3} Benatti F and Floreanini R 2006 {\it J. Phys. A: Math. Gen.} {\bf 39} 2689

\bibitem{per} Peres A 1996 {\it Phys. Rev. Lett.} {\bf 77} 1413

\bibitem{sim}
Simon R 2000 {\it Phys. Rev. Lett.} {\bf 84} 2726

\bibitem{lin}
Lindblad G 1976 {\it Commun. Math. Phys.} {\bf 48} 119

\bibitem{rev}
Isar A, Sandulescu A, Scutaru H, Stefanescu E and Scheid W 1994 {\it Int.
J. Mod. Phys. E} {\bf 3} 635

\bibitem{san}
Sandulescu A, Scutaru H and Scheid W 1987 {\it J. Phys. A: Math. Gen.} {\bf
20} 2121

\bibitem{arus}
Isar A 2007 {\it J. Russ. Laser Res.} {\bf 28} 439

\bibitem{aijqi}
Isar A 2008 {\it Int. J. Quantum Inf.} {\bf 6} 689

\end{thebibliography}
\end{document}